# Multiple-beam Propagation in an Anderson Localized Optical Fiber

Salman Karbasi,[1] Karl W. Koch,[2] and Arash Mafi [1]

[1]Department of Electrical Engineering and Computer Science, University of Wisconsin-Milwaukee, Milwaukee, WI 53211, USA
[2]Optical Physics and Networks Technology, Corning Incorporated, SP-AR-01-2, Sullivan Park, Corning, NY 14831, USA

Abstract: We investigate the simultaneous propagation of multiple beams in a disordered Anderson localized optical fiber. The profiles of each beam fall off exponentially, enabling multiple channels at high-density. We examine the influence of fiber bends on the movement of the beam positions, which we refer to as drift. We investigate the extent of the drift of localized beams induced by macro-bending and show that it is possible to design Anderson localized optical fibers that can be used for practical beam-multiplexing applications.

---

[1]e-mail: `mafi@uwm.edu`

# 1 Introduction

Multicore optical fibers are an increasingly attractive technology for many applications, such as in optical communications [1, 2, 3, 4], sensing [5], optical interconnects [6], optical coherence tomography [7], and imaging [8]. The number and size of waveguiding cores in a multicore fiber depend on the application. Multicore fibers used for imaging or optical interconnects [8, 6, 7, 5, 9] can contain hundreds of cores. On the other hand, multicore optical fibers used for optical-fiber communications are limited to a handful of cores, because the crosstalk between the cores is more detrimental than it is in imaging and interconnect applications; the higher number of cores results in smaller core separation and higher crosstalk and degrades communications.

We recently reported on the development of a novel nano-engineered optical fiber [10], which can support the simultaneous propagation of multiple beams with potential applications in spatially multiplexed optical-fiber communications and imaging. The beam propagation mechanism in this nano-engineered fiber is based on transverse Anderson localization originally proposed by De Raedt et al. [11], and experimentally observed in various configurations [12, 13, 10]. The refractive index profile of the disordered fiber is invariant in the longitudinal direction; however, the transverse index profile is random. We described the fabrication procedure for a polymer version of an Anderson localized optical fiber (p-ALOF) using polystyrene (PS), $n_1 = 1.59$ and poly(methyl methacrylate) (PMMA), $n_2 = 1.49$ in detail in Ref. [10]. Unlike conventional optical fibers that operate on the index-guiding mechanism (total internal reflection), strong multiple scattering across the fiber traps the beam in the transverse direction in the disordered fiber as the beam propagates in the longitudinal direction.

An important difference between a conventional optical fiber and a disordered fiber of Ref. [10] is that the only bound modes in the conventional optical fiber are those confined to the core of the fiber; by contrast, transverse localization guides a beam launched at any point across the transverse profile of a disordered fiber. In an enclosed movie in Ref. [14], we showed that if the incoming beam of light is scanned across the input facet of the fiber, the outgoing beam follows the transverse position of the incoming beam and shifts its location. Here, we propose that this interesting property of the disordered fiber can be used in multiple-beam propagation for spatially multiplexed communication or imaging.

In order for the disordered fiber to be a viable medium for applications that benefit from spatial beam-multiplexing, we need to address two issues inherent in the design of these fibers. First, the localization mechanism is a statistical phenomenon based on multiple random scattering; therefore, the radius of the localized beam in the disordered optical fiber varies from position to position across the fiber. In the design presented in Ref. [10], this variation is approximately 15% of the average beam radius observed in the experiment. Fortunately, this variation can be reduced by increasing the index difference between the random sites of the disordered fiber or by operating at a shorter incident wavelength, as shown in Ref. [14]. The possibility of this reduction is rooted in the self-averaging behavior observed in this random process in the case of strong scattering [10, 16, 15].

A disordered fiber with an index difference of 0.5 between the random sites (air holes in glass) was recently presented in Ref. [17]; however, the air hole density was too low to reduce the variations in the beam radius. Ideally, the air hole density must be near 50% [14] and further optimizations will be required in the future to improve the design.

Second, it is possible that the spatially multiplexed beams drift across the fiber when the fiber is subjected to substantial macro-bending. If the positions of the receivers at the output end of the fiber are initially spatially aligned with the multiple output beams of the ALOF, then the drift of the beams resulting from dynamic macro-bending could result in misalignment and potential loss of the signal. The intention of this paper is to investigate the impact of macro-bending-induced drift in the center of the localized beams and show that it is possible to design Anderson localized optical fibers that can be used for practical beam-multiplexing applications. Our studies are mainly focused on the p-ALOF that was presented in Ref. [10], because it has the smallest localization radius among different samples that we have fabricated so far, and it allows us to compare our theoretical simulations with experimental observations. However, we numerically explore the beam walk-off effect in glass-airhole fibers as well, anticipating future interest in, and development of, these fibers [17].

## 2 Multiple-beam propagation through a disordered fiber

In order to explore the propagation of multiple beams in a disordered fiber numerically, we use a finite difference beam propagation method (FD-BPM), as described in Ref. [14]. We choose the incident wavelength to be 405 nm, which is also the wavelength of the laser used in our experiments. The refractive index distribution used for our numerical simulations is similar to that of the p-ALOF in Ref. [10], where $0.9\mu m \times 0.9\mu m$ sites are assigned refractive index values of $n_1 = 1.59$ or $n_2 = 1.49$ with equal probabilities. In order to observe a multiple-beam propagation effect in a p-ALOF, we launch five incident beams, each with 2.4 $\mu m$ initial beam radius. In Fig. 1(a), we show the intensity profile after 5 cm of propagation along the fiber; the four exterior beams are launched at a distance of 70 $\mu m$ from the central beam. The output beams are observed to remain in the same spatial transverse position across the fiber as launched. In order to confirm our numerical observations, we carried out a similar experiment on a segment of p-ALOF, using the same procedure as in Ref. [10].

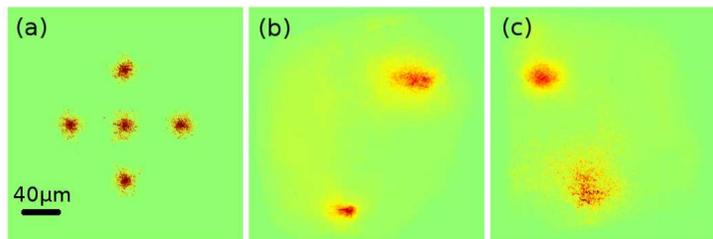

Figure 1: Multiple-beam propagation in a 5 cm-long p-ALOF (a) simulation for five beams; (b) experiment for two beams; and (c) experiment for two beams with different wavelengths. All beams are at 405 nm wavelength, except the bottom-middle beam in subfigure (c), which is at 633 nm wavelength.

In Fig. 1(b), we show the output intensity from the p-ALOF, imaged on a CCD camera beam profiler using a 40x objective. The input double-beam is from two Thorlabs SMF630hp fibers, which are glued alongside each other after their jackets are stripped. We note that the cladding diameter of the SMF630hp fiber is 125 $\mu m$, and we estimate that the two cores were separated by about 190 $\mu m$, after the fibers were glued together.

Each fiber is illuminated separately using a 405 nm diode laser, and the double-fiber setup is butt-coupled to the p-ALOF sample that is 5 cm long. The measured output beam profile clearly illustrates that the two beams can be distinguished across the fiber in the output port.

In Fig. 1(c), we repeat the same experiment of Fig. 1(b), but replace one of the 405 nm diode laser sources with a He-Ne laser operating at 633 nm wavelength. In Ref. [14], we showed that the localized beam radius is larger for longer wavelengths. In Fig 1(c), the bottom-middle beam is at 633 nm wavelength and clearly has a larger localization radius than the top-left beam at 405 nm wavelength.

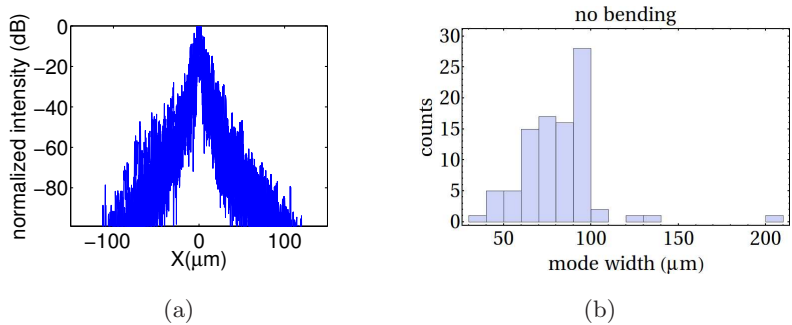

Figure 2: (a) The cross-section of the intensity profiles of the localized beam at 405 nm wavelength for 20 different realizations of the p-ALOF randomness are shown using numerical simulations, where the profiles are plotted on top of each other to capture the expected variations. (b) The experimental measurements of the mode width are shown in a histogram from 92 separate measurements.

We note that the Anderson localization is a statistical phenomenon and the localization beam radius must be calculated by averaging over the elements of the "statistically" identical ensemble of random p-ALOFs. However, the large index difference between the random sites of the p-ALOF results in the "self-averaging" behavior (see, for example,

Refs. [15, 16]); therefore, similar levels of localization are observed for different randomly selected profiles. This self-averaging behavior is essential in ensuring that the p-ALOF presented in here works as a true optical fiber (in the usual sense) and does not need to rely on statistical averaging of multiple samples to localize and guide the optical beam.

We note that despite the strong self-averaging behavior observed in p-ALOFs as also reported earlier in Refs. [10, 14], some level of sample-to-sample variation remains and must be carefully studied, in order to ensure that our observations and conclusions hold well regardless of a specific random realization of the p-ALOF. In Fig. 2(a), we plot the intensity cross-section of the localized beam at 405 nm wavelength for 20 different realizations of the p-ALOF randomness using numerical simulations, where the localized beam intensity profiles are plotted on top of each other to help visualize the expected variations more clearly. We also note that the plot is presented in logarithmic scale to enhance the visual effect of the variation. The observations are in agreement with the previously reported results on the localization of the beam radius and the self-averaging behavior at the wavelength of 405 nm [14].

The numerical simulations in Fig. 2(a) can be compared with the experimental measurements presented in Fig. 2(b). In Fig. 2(b), we show a histogram of the experimental measurements of the mode width from 92 separate measurements. The data is collected by scanning the input beam from a piece of Thorlabs S405-HP fiber over the tip of ten different p-ALOF samples and making nearly nine separate measurements for each fiber sample. We note that the general difference observed between simulation and experiment is consistent with imperfections in preparing the fiber samples (as explained below) and the noise in the CCD beam profiler at low intensities.

We note that in general, there are other parameters, besides the wavelength and the disorder strength, that can affect the variation of the localized beam radius across the fiber [14]. Other than the expected fluctuations due to the statistical nature of the localization, the quality of the fiber surface polishing and the local roughness can play an important role in sample-to-sample and region-to-region variations, especially in polymer-based fibers, where cleaving and polishing are more difficult.

The five-beam intensity profile shown in Fig. 1(a) relates to a single simulation. Although the beams remain well-separated and localized due to the self-averaging behavior for different random realizations of the p-ALOF, we expect that the statistical averaging will reduce both the noise and the overlap between the beams. In Fig. 3(a), we show the five-beam intensity profile after averaging over 20 different simulations from a statistically identical ensemble of p-ALOFs; the averaged individual beams in Fig. 3(a) look considerably cleaner and more circularly symmetric compared with those from the single simulation of Fig. 1(a).

Although the individual beams in Figs. 1(a) and 3(a) look well-separated, this separation needs to be properly quantified, because making judgments solely based on the color scaling in such figures can sometimes be misleading. In order to verify the separation of the beams, we slice the beam intensity profile of Fig. 1(a) along the x-axis at the center (y=0), and plot the cross section of the beam intensities in Fig. 3(b). In fact, in Fig. 3(b), we plot the results of 20 different random simulations on top of each other to show the extent of the possible variations due to randomness. Fig. 3(b) clearly shows that the exponentially decaying tails of the localized beams remain separated to better than 40 dB in intensity, which is beyond the dynamic range of the common CCD cameras. As we showed in Ref. [14], the localization radius of the beam in the p-ALOF used in the present work is about 8 $\mu m$ (based on numerical simulations) with the standard deviation of about 3 $\mu m$ (at 405 nm wavelength); therefore, the beams are separated by 18 standard deviations, which is consistent with an intensity overlap of 40 dB, considering the fluctuations around the average intensity.

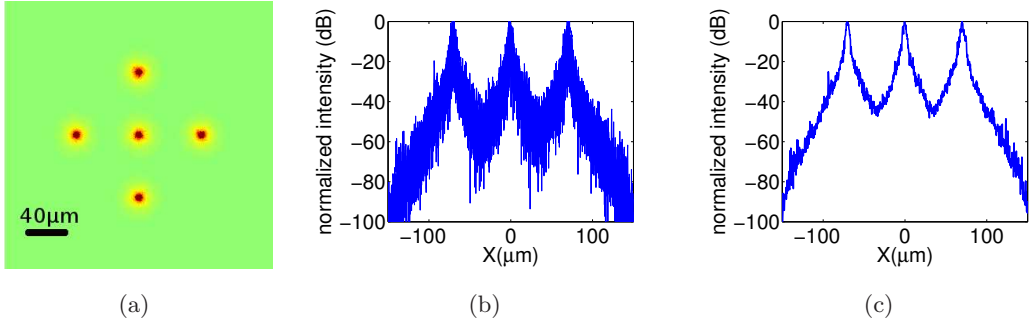

(a) (b) (c)

Figure 3: (a) Similar to Fig. 1(a), but the beam intensity is averaged over 20 different realizations of randomness. Substantial beam clean-up is observed compared with Fig. 1(a) due to the averaging. (b) Cross-section of the intensity profile where the results of 20 different realizations are plotted on top of each other to show the extent to which the beams overlap due to the statistical nature of the problem. (c) Same as (b) but the cross-sectional intensity is plotted for the average of the 20 different realizations. All figures are shown at 405 nm wavelength.

In order to see the beam clean-up due to the averaging process, we take the 20 different random simulations of Fig. 3(b) and show their average in Fig. 3(c) (instead of plotting them on top of each other as we did in Fig. 3(b)). The averaged beam looks substantially cleaner with fewer fluctuations, as expected. We would like to emphasize that in real device realizations, one cannot likely rely on the ensemble averaging; the self-averaging must be strong enough to ensure that the localization and the beam separation can be observed in every element of the ensemble to the desired level. As shown here, statistical simulations are required to capture the degree of fluctuations in order to determine the minimum beam-to-beam separation given the device tolerance for the beam overlap. We note that the short wavelength of 405 nm used here helps in reducing the fluctuations due to strong self-averaging, as already discussed in Ref. [14].

# 3 Impact of macro-bending on the drift of the center of localized beam

As we discussed above, macro-bending can potentially result in a drift in the center of the localized beams in Anderson localized fibers. In order to investigate this drift, we use conformal mapping to model the bending of the optical fiber for our numerical simulations [18]. We assume that the fiber has a refractive profile of $n(x,y)$; when the fiber is bent, it can be conformally mapped to a straight fiber with a modified refractive index profile of $n'(x,y) = n(x,y)\exp(-x/R)$. Here, it is assumed that the p-ALOF is bent in the x-transverse direction with a bend radius of $R$.

We note that according to Ref. [19], the elasto-optical coefficients should be included in the mapped refractive index profile for accurate modeling of the bend. In this work, we choose not to consider the elasto-optical effect; however, this choice will not impact our conclusions in this work.

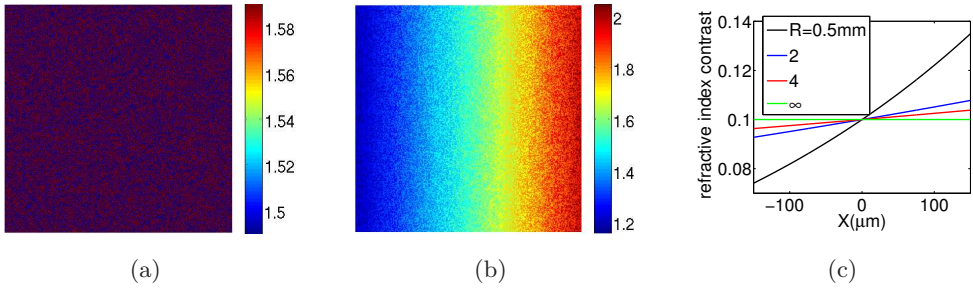

Figure 4: (a) Original index profile of the p-ALOF. (b) Conformally modified refractive index profile of a p-ALOF with bend radius of 0.5 mm. (c) Effective refractive index difference between the low-index and high-index sites for different values of bend radius as a function of the location across the fiber profile. The fiber is assumed to be bent in the x-direction. The dimensions of subfigures (a) and (b) are 300 $\mu m$ on each side.

The original and modified refractive indexes, $n(x,y)$ and $n'(x,y)$ are compared in Figs. 4(a) and (b) for the bend radius of $R = 0.5$ mm, for a sample p-ALOF. The side width of each square region shown in Fig. 4 is $d = 300$ $\mu m$. Fig. 4(b) shows that the $n'(x,y)$ varies considerably in the x-direction due to the bending effect. While the refractive index structure is locally random, both the overall index and the local effective index differences between the random sites have a non-zero gradient due to the bending effect. Fig. 4(c) shows the effective refractive index difference between the low-index and high-index sites (after conformal mapping is included), for different values of bend radius, as a function of the location across the fiber profile.

We can now investigate whether the macro-bending can cause a drift in the location of the center of the localized beam. The center of the beam is defined in Ref. [20] and can shift if the bending effect is stronger than the localization effect. In Fig. 5(a), we show the "transverse" trajectory of the center of an optical beam across the fiber as the beam propagates along the p-ALOF for 5 cm. For the numerical results shown in

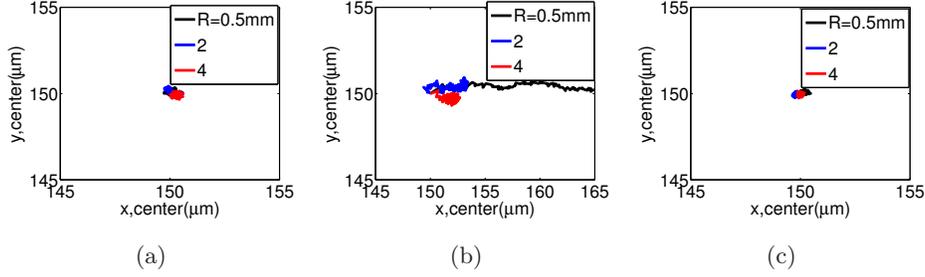

(a)          (b)          (c)

Figure 5: Trajectory of the beam center across the fiber as the beam propagates along a 5 cm segment for different bend radii in a) polymer fiber at $\lambda = 405$ nm, b) polymer fiber at $\lambda = 633$ nm, c) glass fiber at $\lambda = 633$ nm.

Fig. 5(a), the wavelength is 405 nm and the bend radius is $R = 0.5$ mm, $R = 2$ mm, and $R = 4$ mm. The small bend-radius of $R = 0.5$ mm is a worst-possible scenario for a reasonable practical application. For all values of the bend radius, no serious walk-off is observed even after 5 cm of propagation and Anderson localization is observed to dominate over macro-bending.

In Fig. 5(b), we carry out a similar numerical experiment, but at a longer wavelength, 633 nm. As discussed in Ref. [14], transverse Anderson localization is weaker for longer wavelengths. Therefore, in Fig. 5(b) we observe that macro-bending dominates the localization effect, more so for $R = 0.5$ mm (black line) than $R = 2$ mm (blue line) and $R = 4$ mm (red line). For the same wavelength, 633 nm, we expect to see that the transverse Anderson localization will dominate the effects of macro-bending, if the index contrast is raised, as explained in Ref. [14]. In Fig. 5(c), we carry out a numerical experiment similar to that of Fig. 5(b), yet with a larger index difference between the random sites ($n_1 = 1.5$, $n_2 = 1.0$), and observe no serious walk-off in the trajectory of the beam center over 5 cm of propagation, regardless of the bend radius. We note that when there is substantial walk-off, the beam does not preserve its shape and develops a considerable ellipticity in its profile in the direction of the bend [21].

In order to verify our numerical calculations for the drift of the beam center, we carry out an experiment on a sample p-ALOF, where the results are shown in Figs. 6(a) and (b). In Fig. 6(a), we measure the output beam profile of the propagated Anderson localized beam at 405 nm wavelength in a 15 cm-long p-ALOF. We then bend the fiber with a bend radius of approximately 1 mm and measure the beam profile again, as shown in Fig. 6(b). The bend is applied to a 10 cm section of the 15 cm-long p-ALOF, resulting in 16 turns; the remaining 5 cm was left to hold the fiber in the set-up. We observe no noticeable beam walk-off effect, which is also consistent with our numerical simulations in Fig. 5(a).

We note that the numerical results presented in Figs. 5(a), 5(b), 5(c) and the experimental results in Fig. 6 are each for a single realization of the randomness in the fiber without any averaging. In order to confirm that the localization in the bent p-ALOF at

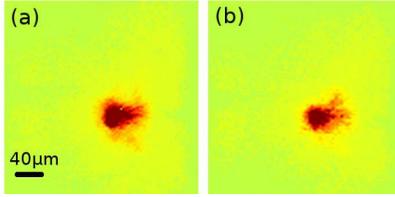

Figure 6: Experimental measurement of the intensity of the propagated light in a fiber with (a) no bend, (b) bend radius of 1 mm. The wavelength is 405 nm and the fiber sample is 15 cm long. No shift is observed, which is also consistent with the simulations in Fig. 5(a). We have intentionally saturated the CCD camera slightly to illustrate the location of the beams with respect to the boundary of the fiber for easier comparison.

405 nm wavelength survives multiple realizations of the random profile, we plot in Fig. 7 a histogram of the experimental measurements of the mode width from 72 separate measurements in bent fibers with the bend radius of approximately 1 mm. We remind that the variations are due to both the statistical variations of the localization phenomenon as well as the imperfections in preparing the fiber samples (such as variations in polishing).

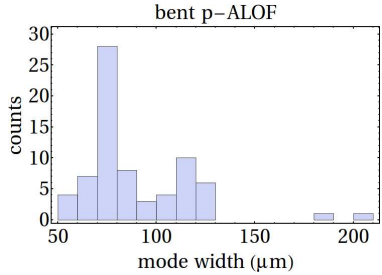

Figure 7: Histogram of the experimental measurements of the mode width from 72 separate measurements in bent fibers with the bend radius of approximately 1 mm. The localization behavior holds for the majority of the 72 random realizations explored in this figure.

We showed in Fig. 4(c) that the effective index difference between random sites varies across the bent fiber. Therefore, we expect the localization effect to be stronger in the region with a higher index difference, i.e., the inside of the bend. In order to observe this effect numerically, we launch three separate beams in a 5 cm fiber bent with a bend radius of 0.5 mm, where one beam is close to the inside of the bend (Fig. 8(a)), another is launched at the center of the fiber (Fig. 8(b)), and the other is close to the outside of the bend (Fig. 8(c)). The output profiles clearly show that the output beam in Fig. 8(c) has the smallest localization radius due to a larger effective index difference between the random sites of the p-ALOF. In Fig. 8(d), we compare the cross section of the averaged intensity for 20 realizations of randomness, where the red profile is related to the less-localized beam near the inside of the bend and the blue profile is related to the more-localized beam near the outside of the bend. We note that in the numerical experiments quoted above, the distance of each beam is 70 $\mu m$ from the center of the fiber, directly across the bending coordinate (which is the x-axis). We note that the relative beam intensity is plotted down to $-100$ dB to clearly show the difference in the localization effect. However, in practice, it may be difficult to experimentally observe this difference between the localization radii, especially if the difference is smaller than or comparable to the variations inherent in the statistical nature of Anderson localization. This is obviously the case for p-ALOF. Moreover, 0.5 mm is smaller than any practically interesting value of the bend radius.

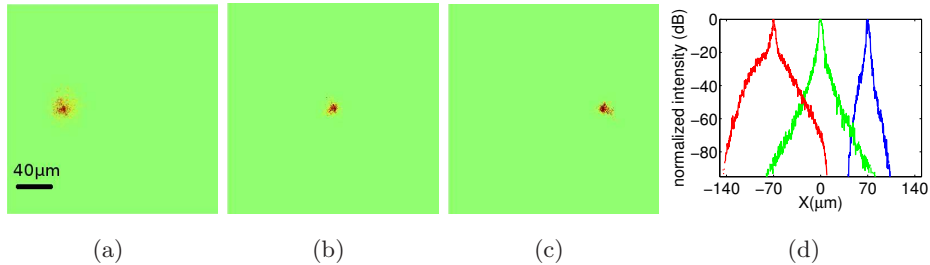

Figure 8: Beam intensity of the propagated light after 5 cm of propagation in a bent p-ALOF with $R = 0.5$ mm, when the light is launched closer to the (a) inside of the bend, (b) center of the fiber, and (c) outside of the bend. (d) Cross section of the beam intensity averaged over 20 samples for the beam in subfigure (a) in red versus the beam in subfigure (b) in green color versus the beam in subfigure (c) in blue color.

## 4 Conclusion

We have shown, both numerically and experimentally, that a p-ALOF can effectively work as a multicore optical fiber. We have also shown that not only can we still observe transverse Anderson localization in a bent disordered fiber, but when the refractive index difference between the random sites is sufficiently large compared with the change in the effective index produced by the bent fiber across the beam, there is also no substantial shift in the center of the beam for reasonable values of the bend radius.

Although averaging over different realizations of the random p-ALOF results in a higher quality of localization, one cannot rely on such ensemble averaging in practical device applications; rather, the self-averaging must be strong enough to ensure that the localization and the beam separation can be observed in every element of the ensemble to the desired level. In the p-ALOFs studied here, the beam overlap was shown to be suppressed to better than 40 dB for the beam-to-beam separation of 70 $\mu m$ using numerical simulations. In the measured samples, the tails of the beams usually extended farther than predicted by theory, most likely because of imperfections in preparing the fiber samples. We also show experimentally that the beam profiles remain well-localized at 405 nm wavelength even when the fibers are bent with a bend radius of around 1 mm. Future efforts are focusing on glass ALOFs with the site-to-site refractive index difference of 0.5 that can result in robust localization of multiple beams at wavelengths longer than 405 nm.

## Acknowledgments

This research is supported by the grant number 1029547 from the National Science Foundation.